\documentclass[reprint, amsmath,amssymb, aps,]{revtex4-1}

\usepackage{graphicx}
\usepackage{dcolumn}
\usepackage{bm}
\usepackage{tabularx}
\usepackage{amsfonts}
\usepackage{hyperref}
\usepackage{amsmath}
\usepackage{leftidx}
\usepackage{braket}
\usepackage{scrextend}
\usepackage[caption=false]{subfig}
\usepackage{color}
\usepackage[dvipsnames]{xcolor}
\usepackage{array}

\hypersetup{backref,pdfpagemode=FullScreen,colorlinks=true,allcolors=blue}
\newcolumntype{L}{>{\raggedright\arraybackslash}X}

\begin{document}

\title{Predicting the Curie temperature of ferromagnets using machine learning}
\author{James Nelson}
\email{janelson@tcd.ie}
\author{Stefano Sanvito}
\email{sanvitos@tcd.ie}
\affiliation{
	School of Physics, 
	AMBER and CRANN Institute, 
	Trinity College, 
	Dublin 2, 
	Ireland
}
\date{\today}

\begin{abstract}
The magnetic properties of a material are determined by a subtle balance between the various interactions at play,
a fact that makes the design of new magnets a daunting task. High-throughput electronic structure theory may help 
to explore the vast chemical space available and offers a design tool to the experimental synthesis. This method 
efficiently predicts the elementary magnetic properties of a compound and its thermodynamical stability, but it is  
blind to information concerning the magnetic critical temperature. Here we introduce a range of machine-learning models to 
predict the Curie temperature, $T_\mathrm{C}$, of ferromagnets. The models are constructed by using experimental data
for about 2,500 known magnets and consider the chemical composition of a compound as the only feature determining
$T_\mathrm{C}$. Thus, we are able to establish a one-to-one relation between the chemical composition and the 
critical temperature. We show that the best model can predict $T_\mathrm{C}$'s with an accuracy of about 50~K. Most
importantly our model is able to extrapolate the predictions to regions of the chemical space, where only a little fraction 
of the data was considered for training. This is demonstrated by tracing the $T_\mathrm{C}$ of binary intermetallic alloys 
along their composition space and for the Al-Co-Fe ternary system.
\end{abstract}

\maketitle

\section{Introduction}

Magnets~\cite{coey2010magnetism,blundell2001magnetism}, compounds in which the atomic spins arrange 
themselves yielding a macroscopic order, are known since antiquity, but still represent a fascinating class of 
materials. In these, the interplay between the local Hund's coupling, the exchange interaction and the magneto-crystalline 
anisotropy, is able to generate a multitude of ground states, which may differ both at the microscopic and macroscopic 
level. Often the particular magnetic configuration of a material is the result of a subtle balance between the interactions 
at play, so that the prediction of the magnetic state based solely on chemical and structural information is a delicate 
exercise. Probably the largest subset of magnetic compounds is populated by ferromagnets, where the atomic spins align 
along the same direction. Regardless of the specific magnetic phase, a magnet loses its collective order at the critical 
temperature that, in the case of a ferromagnet, is known as the Curie temperature, $T_\mathrm{C}$. This means that at 
and above $T_\mathrm{C}$ a ferromagnet ceases to be magnetic. 

\begin{figure}[h]
	\centering\includegraphics[width=\columnwidth]{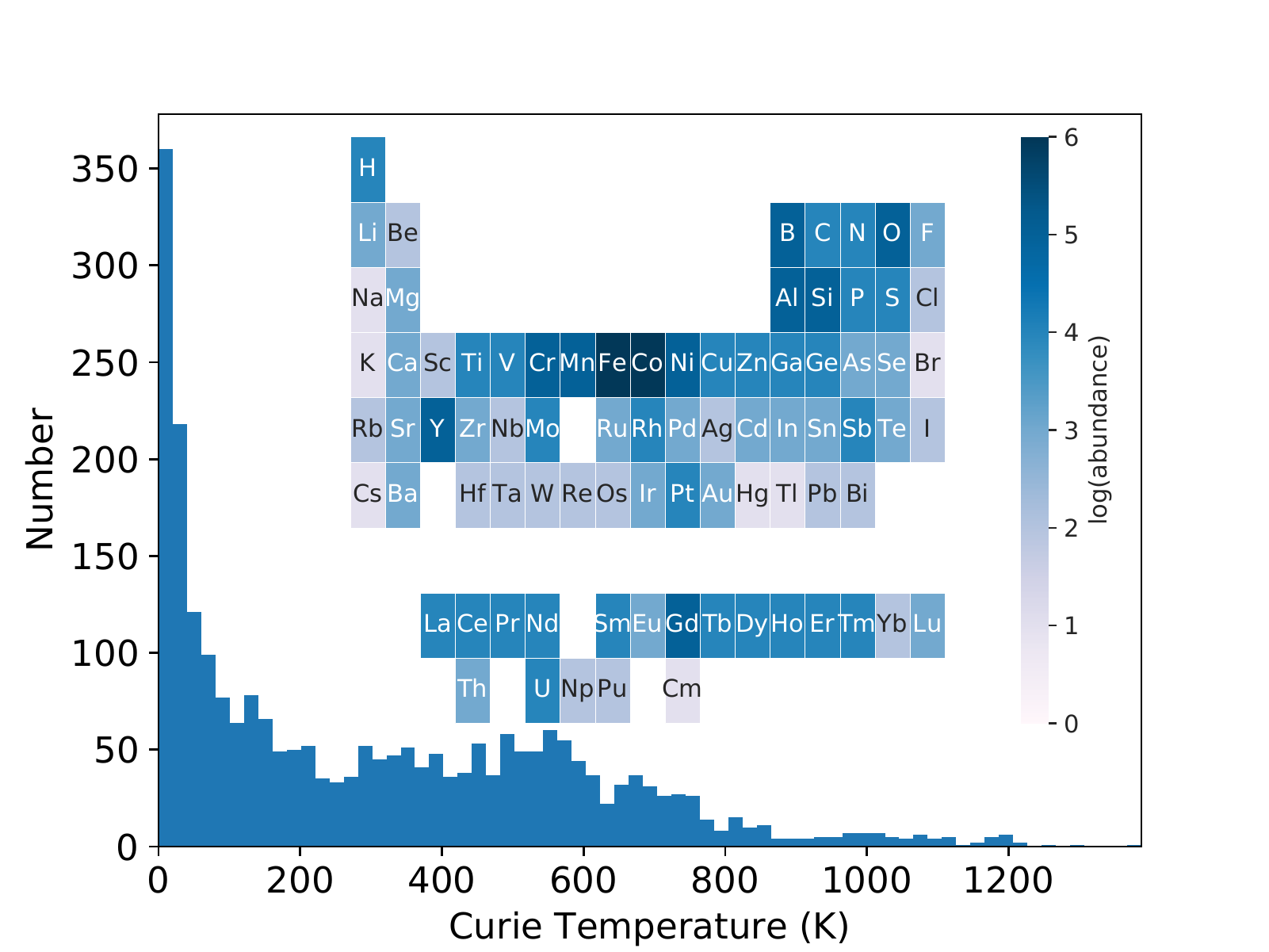}
	\caption{Histogram of the $T_\mathrm{C}$'s of about 2,000 known ferromagnets. The median value 
	of the distribution is 227~K. The insert shows the relative elemental abundance, in logarithmic scale, 
	for the ferromagnets included in the dataset (the logarithm of the number of compounds containing 
	a particular element). The most frequent magnetic element is Co followed by Fe and Gd.
}
	\label{fig:cuire_bar}
\end{figure}

When a magnet is then employed in a given technology, for instance in energy production and transformation 
or in data storage, its $T_\mathrm{C}$ must significantly exceed room temperature. This means that typically 
a magnet will be considered as `useful', if its Curie temperature is around 600~K. Unfortunately, not many 
magnetic compounds reach such value. In Fig.~\ref{fig:cuire_bar} we present the distribution of the measured 
$T_\mathrm{C}$'s of about 2,500 known ferromagnets (see later for details). The median of the distribution is 
227~K, meaning that the vast majority of ferromagnets known to date are actually paramagnetic at room temperature. 
Furthermore, it is clear that the number of compounds satisfying the $T_\mathrm{C}>600$~K criterion is only a small 
fraction of the total, suggesting that finding new `useful' ferromagnets is indeed a rare event and welcome news.

Figure~\ref{fig:cuire_bar} also presents the relative elemental abundance for the ferromagnets 
included in the dataset, namely for every element the number of compounds that contain that 
given element. As expected the vast majority of the ferromagnets contains at least one of the 3$d$ 
magnetic transition metals, Fe, Co, Ni and Mn, with Al being the most frequent of the non-magnetic 
ions. However, it is interesting to note that, with the only exception of noble gases and highly radioactive 
elements, magnets can be made by incorporating essentially any ion in the periodic table. This gives us 
a potentially very large chemical space to explore when designing new magnets. 

In the last few years there have been a few attempts at systematically predicting the existence of new 
magnets ahead of experiments. These study are typically based on high-throughput numerical 
explorations~\cite{CurtaReview}, where the electronic structure of hypothetical compounds is computed 
at the level of density functional theory. The construction of the prototypes to calculate usually proceeds by 
substituting elements in known phases~\cite{Elsasser2018a,Elsasser2018b}, or by exploring the entire chemical 
space compatible with a given crystal structure~\cite{SciAdv2017}. In both cases one can extract some 
elementary magnetic properties of the prototypes (magnetic moment per cell, density of states, 
magneto-crystalline anisotropy, etc.) and possibly assess their thermodynamical stability, namely one can 
forecast if a given prototype can be made and what its magnetic properties would be. 

However, little information about $T_\mathrm{C}$ can be extracted, unless the specific class of compounds 
investigated satisfies some simple empirical rules. For example the Curie temperature of Heusler alloys of 
composition Co$_2XY$, with $X$ and $Y$ being either a transition metal or a main group element, follows a 
Slater-Pauling curve~\cite{Felser2011}, while Mn-containing magnets can be arranged along the phenomenological 
Castelliz-Kanomata curves~\cite{Castelliz,Kanomata}. In the absence of such empirical rules the search for 
new compounds remains blind to $T_\mathrm{C}$. This is a rather severe deficiency, since the calculations 
cannot distinguish from the outset which regions of the chemical space may yield high-$T_\mathrm{C}$ magnets. 
As a consequence most of the discovery computational effort is usually spent for materials with little technological 
potential. 

Note that the prediction of $T_\mathrm{C}$ by first-principle methods is not an easy task either. The most
common strategy consists in mapping electronic structure calculations onto some effective Hamiltonian, most 
typically the Heisenberg model, which is then used to extract $T_\mathrm{C}$ via Monte Carlo techniques. 
This is a valid approach only if the relevant part of the magnetic excitation spectrum has a spin-wave nature,
which is not universally guaranteed. As such the same method applied to different materials may result in
predictions of $T_\mathrm{C}$'s with a very different level of agreement to experiments~\cite{Turek2006}.
Certainly, the computation of the entire magnetic excitation spectrum with density functional theory is possible
\cite{PhysRevLett.102.247206}, but the computational overheads are significant. Furthermore, also in this 
case uncertainty and errors arise from the choice of the exchange and correlation functional and from the other
approximations taken. Overall, first-principle methods are very valuable for understanding the origin of
a particular $T_\mathrm{C}$, but they are currently not fit to be used as a prediction tool ahead of experiments.
For all these reasons it would be desirable to construct a universal predictor for $T_\mathrm{C}$ based solely on 
chemical information. 

The present work responds to this demand. We introduce a range of machine-learning (ML) models, uniquely based
on known experimental data for about 2,500 ferromagnets, which can predict the Curie temperature using only 
chemical information. Machine learning is rapidly becoming a valuable tool in physics and materials science and 
it has been used already for a relatively wide range of problems. These go from the accelerated discovery of new 
materials~\cite{pilania2013accelerating,meredig2014combinatorial}, to the estimation of materials physical 
quantities~\cite{rupp2012fast,Isayev2017}, to the definition of novel density functionals~\cite{snyder2012finding, James2019}. 
Here we use ML to unveil patterns in the chemical-to-$T_\mathrm{C}$ relation, and use these patterns as 
predictors for the $T_\mathrm{C}$ of new materials. Our approach is in the same spirit of that used by 
Stanev and co-workers~\cite{stanev2018machine} to sort superconductors according to their critical 
temperature.

Our paper is structured as follows. Firstly, we introduce the computational methods used. In particular we discuss
extensively the issues of representing the chemical composition of a given compound, of creating and processing the 
dataset, and of training and choosing the most appropriate and best performing ML model. We then discuss the performance 
of the various models against our dataset and show their ability to extrapolate in regions of data, where little
information were available during the model training. These, for instance, include the prediction of $T_\mathrm{C}$
along the composition space of different transition-metal binary systems and for the Al-Co-Fe ternary one. Finally 
we outline some possible ways to include information about the materials structure in the ML model and 
its effect on the overall accuracy.

\section{Methods}

\subsection{Definition of the feature vectors}
The goal of supervised ML is to estimate a function, $f$, that maps a feature vector, $\textbf{x}$, to a target 
variable, $y$, namely to find the function $f: \textbf{x} \rightarrow y$ that better interpolates a set of available 
data (note that the target variable does not need to be a scalar quantity). In our case the feature vector 
contains information concerning a given compound, while the target variable is $T_\mathrm{C}$. Although
there are no rigorous stringent rules on how to construct the feature vector, ultimately the success of any ML
strategy resides on the ability to define the most relevant features correlating to a given property. 
When the ML strategy is applied to materials science a number of desirable conditions defining the feature 
vector emerge naturally~\cite{ghiringhelli2015big, jain2016new, faber2015crystal, rupp2012fast, 
von2015fourier}. In particular it was proposed that the feature vector
\begin{enumerate}
	\item should be computationally inexpensive to create.
	\item should be continuous, reflecting the fact that the properties of a material vary continuously. 
	\item should be independent of both the choice of unit cell and of the number of atoms in the unit 
	cell (for periodic solids).
	\item should be able to distinguish different materials with different properties, namely different 
	compounds with different $T_\mathrm{C}$ should have different feature vectors. 
\end{enumerate}
We begin by defining feature vectors that do not include any structural information of compounds, but are
constructed to satisfy the first three conditions. Clearly the last one will be automatically violated, since 
polymorphs of a given chemical composition presenting different $T_\mathrm{C}$'s will be described by 
identical feature vectors. In particular our goal is for the feature vectors to capture the 
chemical composition of a given compound, namely its chemical formula. We can then divide the 
features defining a given compound into three categories: features which only take elemental properties 
into account (e.g. the atomic number), features which take only the compound stoichiometry into account, 
and features which take both into account.

Notably features belonging to the first category can be found to violate the second criterion and, for this reason,
they have been excluded. Take as an example the maximum atomic number, $Z^\mathrm{max}$, 
of a given compound~\cite{ward2016general} and consider the binary alloy $\text{A}_{1-x}\text{B}_x$, 
with $Z_\mathrm{B}>Z_\mathrm{A}$ as an example. At $x=0$ a feature vector containing this component 
will be discontinuous as the Euclidean distance between $\text{A}_1$ and any $\text{A}_{1-x}\text{B}_x$ will 
always be greater than $|Z_\text{A}-Z_\text{B}|$ and will not go to zero smoothly as $x$ goes to zero. In 
order to avoid this shortcoming
we need to consider features that, either explicitly or implicitly, preserve information related to the atomic 
fractions of the various elements in a compound, and exclude those that just relate to the elemental properties.
As such, features like the number of different elements in the compound (2 for binaries, 3 for ternaries, etc.)
or the maximum atomic number are ruled out. In contrast a feature like the mode of the atomic number~\cite{note1} 
is acceptable, since it encodes information about the atomic fractions. 

Possible features created solely from the stoichiometry of a compound include the $L^p$ stoichiometry norm 
\cite{ward2016general} and the stoichiometry entropy. The first one is generally defined as $||x||_p=(\sum_i|x_i|^p)^{1/p}$, 
while the second is $S=-\sum_ix_i\log(x_i)$, where $x_i$ is the atomic fraction of the i-$th$ element. Features based 
on the product of the atomic fractions or on the number of different elements in the compound are ruled out by the 
second principle, using arguments similar to the one given above. In any case, one has still to establish whether 
stoichiometry-only features are informative enough for predicting most quantities.

Finally there are generally two ways to define features that take both elemental properties and stoichiometry into account. One 
possibility is to associate to each compound a high-dimension vector defined as
\begin{equation}
\mathbf{v}_\mathrm{chem}=\{x_\mathrm{H}, x_\mathrm{He}, x_\mathrm{Li}, x_\mathrm{Be}, ...  \}
\end{equation}
where $x_\alpha$ is the atomic fraction of element $\alpha$. For instance, $\text{Fe}_3\text{O}_4$ is represented as 
the vector $(...,0,4/7,0,...,0,3/7,0...)$, with 4/7 assigned at the eighth position (oxygen) and 3/7 to the 26th one (iron).
The advantage of $\mathbf{v}_\mathrm{chem}$ is that it uniquely represents a chemical formula, and its disadvantage
that the vectors tend to be very sparse and high dimensional, hence difficult to train. A second option was suggested 
by Ward et al. \cite{ward2016general} and consists in using composition-weighted elemental-properties. An example of 
these is the composition-weighted atomic number, defined as $\langle Z\rangle=\sum_iZ_ix_i$, with $Z_i$ and $x_i$ being 
the atomic number and the atomic fraction of the element $i$. In the case of $\text{Fe}_3\text{O}_4$ this is $8(4/7)+26(3/7)=15.71$.
Similar quantities can be constructed with analogous definitions. Furthermore, for any elemental quantity $Q$ one 
can also define a composition-weighted mode $\langle |Q|\rangle$, which takes the $Q$ value of the element with the 
highest atomic fraction in the compound and an average in the case of multiple modes, and a composition-weighted 
absolute deviation $\langle \Delta Q\rangle=\sum_i|Q_i-\langle Q\rangle|x_i$. Table~\ref{table:list_of_features} summarises 
the features used in this work.
\begin{table}[h]
	\centering
	\begin{tabular}{|l|c|c|} \hline
		\textbf{Features} & Symbol & Dim. \\
		\hline
		$L^p$ stoichiometry norm ($p=1, 2, 3$) & $||x||_p$ & 3 \\
		Stoichiometry entropy & $S$ & 1 \\
		Atomic fraction vector & $\mathbf{v}_\mathrm{chem}$ & 84 \\
		CW atomic number  & $\langle Z\rangle$,  $\langle |Z|\rangle$, $\langle \Delta Z\rangle$ & 3\\
		CW valence electrons & $\langle N_\mathrm{V}\rangle$,  $\langle |N_\mathrm{V}|\rangle$, $\langle \Delta N_\mathrm{V}\rangle$ & 3 \\
		CW period & $\langle P\rangle$,  $\langle |P|\rangle$, $\langle \Delta P\rangle$ & 3 \\
		CW group & $\langle G\rangle$,  $\langle |G|\rangle$, $\langle \Delta G\rangle$ & 3 \\
		CW molar volume & $\langle V\rangle$,  $\langle |V|\rangle$, $\langle \Delta V\rangle$ & 3 \\
		CW melting $T$ & $\langle T_\mathrm{M}\rangle$,  $\langle |T_\mathrm{M}|\rangle$, $\langle \Delta T_\mathrm{M}\rangle$ & 3 \\
		CW electronegativity & $\langle \epsilon\rangle$,  $\langle |\epsilon|\rangle$, $\langle \Delta \epsilon\rangle$ & 3 \\ \hline
	\end{tabular}
	\caption{The full list of features used in this work. In the last column we report the dimension of any given feature.
	The atomic fraction vector has dimension 84, as not all elements in the periodic table can be found in the ferromagnets
	included in our database (e.g. He, Ar, etc.). In total our feature vector has a dimension of 129. Here
	``CW'' denotes ``composition-weighted''. The numerical values of the elemental properties are taken from Ref.~\cite{ong2013python}.}
	\label{table:list_of_features}
\end{table}

\subsection{Machine-learning models training}
In an ideal situation, where there is abundance of data, the training of ML models proceeds by 
splitting the dataset into three mutually exclusive partitions, a training set, a validation set and a test 
set~\cite{friedman2009elements}. The training set is used to train the model. Most ML models are 
defined by one or more parameters, called hyperparameters, which cannot be learnt from the 
data and thus need to be specified from the outset. The validation set is then used for determining 
the best hyperparameters of any given model and for choosing the best overall model. Finally, the test 
set is used for estimating the generalization error of the model, namely for assessing how well the 
given model performs on never-seen-before data. However, in situations where the datasets are small, 
splitting the data into three sets makes each one of them too small, diminishing the overall ability of the 
model to learn and hence its accuracy. Therefore, in this case the training and the validation sets are combined 
into a single set, which we will refer to henceforth as the training set. Then $K$-fold cross-validation is used to 
determine the hyperparameters and to select the best model. Here the training set is split into $K$ subsets (in this 
work $K$ was chosen to be $K=3$). For each given set the model is trained over the other $K-1$ sets and tested 
over the remaining one, the cross-validation error is the average of all the errors. Then the best model, 
namely the model with the lowest cross-validation error, is trained over the entire training set. Finally, the test 
set is used to estimate the accuracy of the chosen model. 

In order to quantify the model's performance we have used the coefficient of determination, $R^2$. Given a set 
of target variables $\{y^{i}\}$, with mean $\mu$ and predicted values $\{f(\textbf{x}^{i})\}$, $R^2$ is given by
\begin{equation}\label{r2}
R^2 =  1 - \frac{\sum_i[y^{i}-f(\textbf{x}^{i})]^2}{\sum_i[y^{i}-\mu]^2}\:.
\end{equation}
A perfect predictor of the target variables would always score $R^2=1$ on any set 
[$f(\textbf{x}^{i})=y^i$ for $\forall i$]. 

\subsection{Construction of the dataset}
The dataset of experimental $T_\mathrm{C}$ has been constructed by aggregating the following
sources: the AtomWork database \cite{xu2011inorganic}, Springer Materials \cite{connolly2012bibliography}, 
the Handbook of Magnetic Materials \cite{HMM} and the book {\it Magnetism and Magnetic 
Materials} \cite{coey2010magnetism}. A few additional values have been taken from the 
references~\cite{albert1971magnetic, givord1971proprietes, okamoto1989mn}. For a number
of compounds the various databases report multiple values of $T_\mathrm{C}$, and there are
also compounds where the same database returns a range of $T_\mathrm{C}$'s for the same ferromagnet.
Notably in the vast majority of cases the spread of $T_\mathrm{C}$'s about the mean is rather small, 
so that the choice of a particular value is irrelevant. In particular, for 79\% of the compounds associated to
multiple $T_\mathrm{C}$'s the difference between the maximum and minimum value is less than 50~K.
However, there is also a number of compounds presenting a much larger range of reported critical 
temperatures, namely for 4.9\% of the multiple-$T_\mathrm{C}$ data the difference between the maximum 
and minimum $T_\mathrm{C}$ value is greater than 300~K.

There are several reason for these occurrences. In some compounds magnetism is subtly related to the
quality of the sample, so that experiments performed by different groups may report different $T_\mathrm{C}$.
This is particularly relevant, since the various data sources contain $T_\mathrm{C}$'s extracted over different
periods of time, so that the spread of data sometime reflects the improvements in crystal growth over time. 
A second reason is related to polymorphism, namely to the existence of compounds with the same stoichiometry 
but different crystal structure, and hence different $T_\mathrm{C}$. Finally, there are several transition-metal/rare-earth 
intermetallic magnets for which multiple $T_\mathrm{C}$ are reported. For instance the two values of 
186~K~\cite{Carfagna1968} and 641~K~\cite{Buschow1977} have been both reported for $\text{Sm}_2\text{Ni}_{17}$. 
Such large discrepancy has been attributed to the presence of possible secondary phases~\cite{Buschow1977}.
Since we are aiming at constructing ML models that use only chemical information to define compounds,
our feature vector can not include any attribute related to sample quality or polymorphism. As such, in the 
presence of multiple $T_\mathrm{C}$'s we have to establish a criterion to select a single value
for any stoichiometry. In this work we have used the median of the distribution instead of the mean or 
the maximum value as this is more resistant to outliers. For instance, if for a given compound the reported 
$T_\mathrm{C}$'s are 300~K, 305~K and 700~K, their mean is 435~K while their median is 305~K. The
first one is not associated to any real measurement while the second is.

In preparing our dataset we have carefully checked that there is enough diversity in the entries.
Consider the following thought experiment. Suppose that for every entry in our database there
are also many similar entries, namely there are several compounds with similar stoichiometry 
(feature vector) and similar $T_\mathrm{C}$. This may be, for instance, the case of a binary alloy
$A_xB_{1-x}$, where many data are available in a narrow range of compositions. For a dataset
of such homogeneous composition it is likely that almost any ML model will perform well.
However, the same model will be unlikely to perform well for compounds significantly different 
to the ones found in the training set, namely the model will have little ability to generalize to a 
broader composition range. This is, of course, an unwanted feature. 
Our strategy to curate the database is then the following. Firstly, we standardise the chemical
formula notation, by replacing fractional stoichiometry with integer one (e.g. Cu$_{0.5}$Ni$_{0.5}$
becomes CuNi = Cu$_1$Ni$_1$), and by simplify the stoichiometry when possible 
(e.g. Ni$_{75}$Al$_{25}$ becomes Ni$_{3}$Al = Ni$_{3}$Al$_1$). At this point we check for
duplicates and, if these result in multiple $T_\mathrm{C}$'s for the same stoichiometry we take
the median value. Next the dataset is ordered according to the number of atoms in the chemical 
formula and we compute the $L^1$-norm of the atomic fraction vector, $\mathbf{v}_\mathrm{chem}$, between all the 
entries in the database. If the distance between two entries is less than 0.01 we remove the compound 
with the larger number of atoms in its chemical formula. The rational for doing this reflects our intuition 
that Fe$_5$Sn$_3$ and Fe$_8$Sn$_5$, with a distance of 0.019, should be considered 
as two different compounds, while Co$_5$Tb$_1$ and Co$_{5.1}$Tb$_1$, with a distance of 0.005, 
are essentially the same compound. In this last case we keep only the simpler Co$_5$Tb$_1$.

Finally our data needs to enable the ML models to distinguish between magnetic and non-magnetic
materials. This is not an issue if our only goal is that of describing the $T_\mathrm{C}$ of known 
ferromagnets, but it becomes one when the ML model is used as $T_\mathrm{C}$ predictor for
unknown, hypothetical, compounds. As constructed, our data distribution, $p_{\text{data}}$, only 
includes ferromagnets, so that a ML model will not be able to learn from $p_{\text{data}}$ about 
stoichiometries having $T_\mathrm{C}=0$. A possible way out, as suggested by Stanev et 
al.~\cite{stanev2018machine}, is that of first training a classifier to predict whether or not a given compound 
has a $T_\mathrm{C}$ greater than some critical value, $T_{\text{critical}}$. If one sets $T_{\text{critical}}$
relatively low, the classifier will distinguish between magnetic compounds and magnetic only at very low 
temperature. Here, however, we take a somewhat more straightforward approach and we simply include in our 
dataset some non-magnetic materials. In particular we include the elemental phases of all the non-magnetic 
elements of the periodic table. This procedure effectively provides to the ML model some information about 
non-ferromagnets and hopefully it makes it more robust when making new predictions.

After the data processing described above our training set contains 1,866 entries, while the test set 
has 767. About 3\% of the compounds are unaries, including the non-ferromagnetic ones. The 
ferromagnetic elemental phases are: Co ($T_\mathrm{C}=1380$~K), Fe (1040~K), Ni (630~K),
 Gd (290~K), Tb (220~K), Dy (85~K), Nd (30~K), Tm (30~K), Er (20~K), Ho (20~K) and Pr (8.7~K). Binaries and ternaries make up 31\% and 49\% of the dataset respectively with the rest of the dataset having more than 3 distinct elements. 

\section{Results}
\subsection{Machine-learning model performance}
In the construction of the ML models we compare four different algorithms, namely Ridge Regression (Ridge), 
Neural Network (NN), Kernel Ridge Regression (KRR) and Random Forests (RF). The NN is implemented employing
Keras~\cite{chollet2015keras}, while for the rest we use Scikit-Learn \cite{scikit-learn}. For both Ridge and KRR the 
cross-validation set is used to determine the optimum regularization parameter. In contrast the dropout rate
is the hyper-parameter used for the NN, while the maximum depth of the tree is that for RF.
The accuracy of many algorithms can be sometimes improved by reducing the dimension of the feature space. 
This is particularly true when there is correlation between the various features and for ML algorithms affected
by the curse of dimensionality~\cite{bellman1957dynamic}. For a feature vector of dimension $p$, one can 
construct $2^p$ possible subsets of the features, which in our case translates in $\sim10^{39}$ subsets~\cite{Note2}.
It is, therefore, infeasible to perform an exhaustive search for the best performing subset. Instead, we 
have decided to use two different methods of feature-dimension reduction, namely Correlation (C) and 
Principle Component Analysis (PCA), and used such methods together with all the chosen ML algorithms. 
Correlation ranks the features according to the absolute value of the Pearson correlation coefficient, which 
is defined as $c_\mathrm{P}=\text{cov}(x,y)/\sigma_x\sigma_y$, with cov$(x,y)$ being the $(x, y)$ covariance 
and $\sigma$ the standard deviation. In our case $y$ is the $T_\mathrm{C}$ and $x$ corresponds to each 
feature. In contrast, the PCA projects the feature space into a lower dimensional one, while 
trying to preserve as much data variance as possible~\cite{goodfellow2016deep}.  

The results of the 3-fold cross validation score for the different algorithms operated together with different 
feature reduction techniques are shown in Table~\ref{table:1}.
\begin{table}[h]
	\centering
	\begin{tabular}{l|lllllllll}
		$R^2$ & All  & C10 & C20 & C40 & C80 & P10 & P20 & P40 & P80 \\
		\hline
		Ridge & 0.53 & 0.48     & 0.5     & 0.52     & 0.52     & 0.24     & 0.27     & 0.31     & 0.39     \\
		KRR   & 0.69 & 0.72     & 0.72     & 0.72     & 0.69     & 0.69     & 0.72     & 0.70     & 0.68     \\
		NN    & 0.76 & 0.72     & 0.76     & 0.77     & 0.78     & 0.73     & 0.77     & 0.77     & 0.77     \\
		RF    & 0.81 & 0.76     & 0.77     & 0.78     & 0.79     & 0.72     & 0.74     & 0.73     & 0.72    
	\end{tabular}
\caption{3-fold cross-validation $R^2$ score of all the algorithms chosen combined with the various feature reduction 
techniques. ``All'' indicates the case where no feature reduction is applied. ``C'' means Correlation feature reduction 
and ``P'' is for PCA. The number beside the type of feature reduction scheme indicates the size of the reduced feature 
space. Thus, for example, P40 means that PCA has generated a 40-dimensional feature space.}
\label{table:1}
\end{table}
In general we find that the performance of the different algorithms ranks them in the following order Ridge, KRR, NN and 
RF. Dimensionality reduction does not seem to significantly improve the $R^2$ and in fact, with the only exception 
of the Correlation scheme applied to KRR and NN, it appears always better to run the ML algorithm over the full
feature space. Overall Random Forests using the entire set of features performs the best on the cross-validation 
sets and, therefore, it is chosen as the final model. RF achieves a cross-validation $R^2$ of 0.81 and a test one of 
0.87, demonstrating that the cross-validation score is a good estimate of the test error.

\begin{figure}[h!]
	\centering
	\includegraphics[width=\columnwidth]{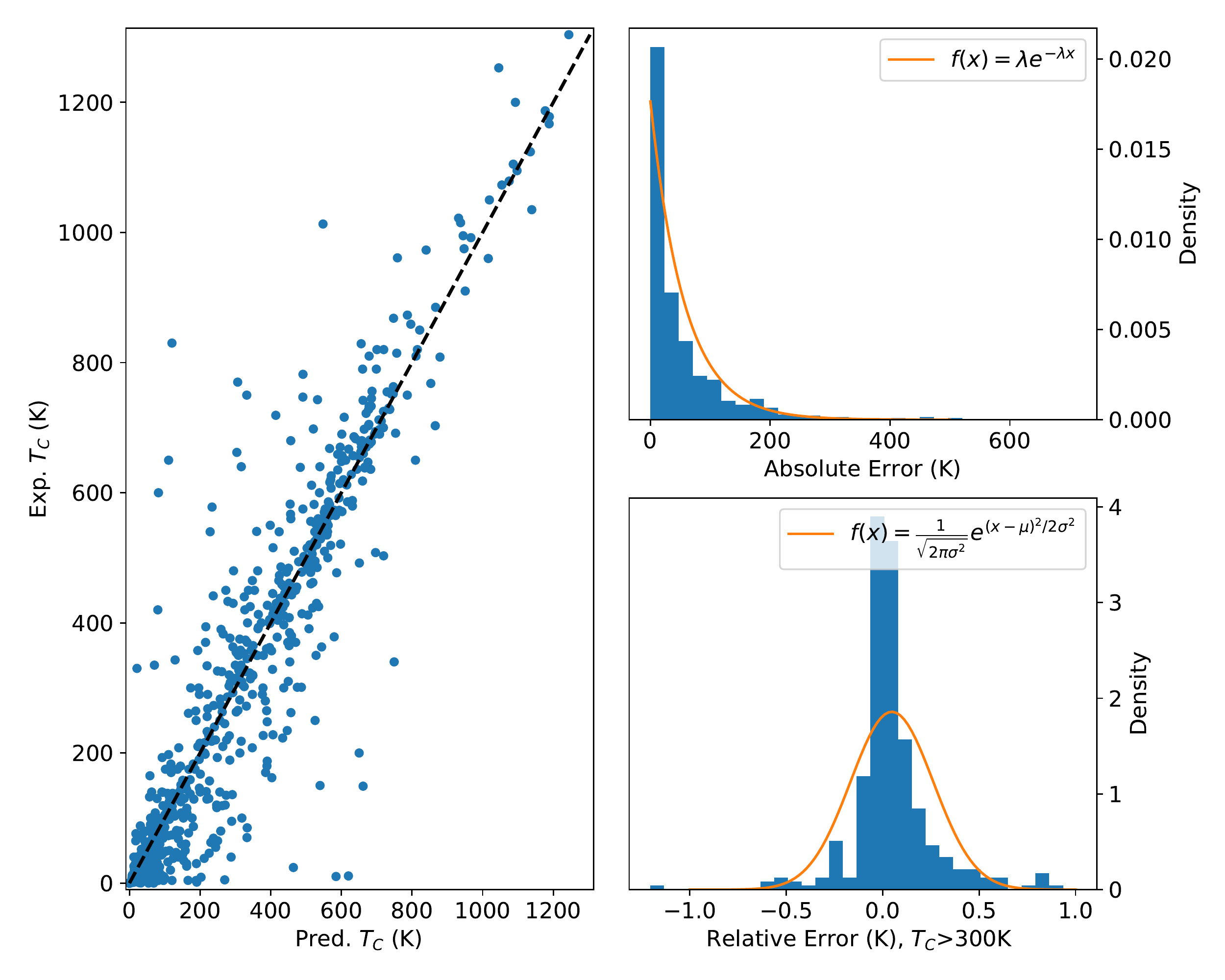}
	\caption{RF model for the $T_\mathrm{C}$ of ferromagnets. In the left-hand side panel we compare the 
	experimental and predicted $T_\mathrm{C}$ for the test set. The $R^2$ coefficient of the model is 0.87 
	and the mean absolute value is 57~K. 
	The upper left panel shows the distribution of absolute errors, 
	$|T_\mathrm{C}^\mathrm{exp}-T_\mathrm{C}^\mathrm{predic}|$, while the lower one displays
	the distribution of the relative errors for compounds with $T_\mathrm{C}>300$~K. The orange lines
	indicate the best fit respectively to an exponential and a Gaussian distribution.}
	\label{fig:tc_test}
\end{figure}

In Fig.~\ref{fig:tc_test} we present our best result for the RF algorithm. Here we plot the predicted $T_\mathrm{C}$'s
against the experimental ones for all the ferromagnets contained in the test set. In addition we present a distribution of
the absolute errors and one for the relative error of compounds presenting $T_\mathrm{C}>300$~K.
In general we find that our ML model can predict relatively well the experimental $T_\mathrm{C}$, in particular 
for Curie temperatures exceeding 300~K. This is important since in this range one finds the magnets useful for 
room-temperature applications. The mean absolute error over the entire distribution is 57~K. Such a value gives
us confidence that the ML model can distinguish between high-$T_\mathrm{C}$ ferromagnets and  
low-$T_\mathrm{C}$ ones, namely it allows us to identify the potential of an hypothetical chemical composition 
against ferromagnetism.

The distribution of the $T_\mathrm{C}$ absolute errors for our best ML model is exponential with decay 
coefficient, $\lambda$, of 0.018. From the distribution one can learn that an $1-e^{-\lambda x}$ fraction 
of the data are with $x$~K of the experimental data. For example 59~\% of the predicted $T_\mathrm{C}$'s 
are within 50~K from the measured ones, and  83~\% are within 100~K. Large absolute errors are found only 
for compounds presenting a rather small $T_\mathrm{C}$, which are erroneously predicted to be robust 
ferromagnets. A more detailed understanding of the performance of our ML model can be obtained by looking 
at the distribution of the relative error, defined as 
$(T_\mathrm{C}^\mathrm{exp}-T_\mathrm{C}^\mathrm{predic})/T_\mathrm{C}^\mathrm{exp}$,
where $T_\mathrm{C}^\mathrm{exp}$ and $T_\mathrm{C}^\mathrm{predic}$ are respectively the average
experimental $T_\mathrm{C}$ and the predicted one. In this case we present data only for compounds with 
$T_\mathrm{C}$ exceeding 300~K (right-hand side panel of Fig.~\ref{fig:tc_test}). There are two main reasons behind
such choice. On the one hand, these are the compounds potentially useful for room-temperature applications. On the
other hand, for ferromagnets presenting low measured $T_\mathrm{C}$, relative small absolute errors may result
in rather large relative ones. In this case the distribution appears symmetric around zero, indicating that our
ML model has no systematic bias towards either overestimating or underestimating $T_\mathrm{C}$. The shape
of the distribution is Gaussian-type with a half-height width of 0.51. We find that only 
5~\% of the compounds present a relative error larger than 50~\%, and only 15~\% have errors larger than
25~\%.

Finally we take a look at whether or not our ML model tends to systematically fail for some particular chemical 
compositions. Again we analyse data only for compounds with $T_\mathrm{C}>300$~K. In Table~\ref{table:2}
we present the elemental abundance, $f_M^\alpha$, of the most relevant magnetic transition metals, $M$, among 
compounds presenting a relative error beyond a give threshold, $\alpha$. For instance, in the cell corresponding 
to $f^{50}_M$ and Cr we show the relative abundance of the Cr element among the compounds 
presenting a predicted $T_\mathrm{C}$ with a relative error larger than 50~\%, $f^{50}_\mathrm{Cr}$. The abundance
is calculated as the total number of compounds presenting a given element divided by the total number of compounds.
Note that the elemental abundances do not necessarily sum up to unity, since a compound may contain more than 
one transition metal.
\begin{table}[h]
	\centering
	\begin{tabular}{l|lllllll}
		{\bf Element} & $f_M^{50}$  & $f_M^{30}$ & $f_M^{20}$ & $f_M^{10}$ & $f_M^{5}$ & $f_M^{2}$ & $f_M^{0}$ \\
		\hline
		Cr & 0.0 & 0.01 & 0.02 & 0.02 & 0.04 & 0.05 & 0.05 \\ 
		Mn & 0.02 & 0.04 & 0.08 & 0.09 & 0.12 & 0.14 & 0.15 \\ 
		Fe & 0.03 & 0.07 & 0.13 & 0.24 & 0.38 & 0.56 & 0.74 \\ 
		Co & 0.01 & 0.02 & 0.03 & 0.06 & 0.11 & 0.17 & 0.21 \\ 
		Ni & 0.01 & 0.02 & 0.02 & 0.03 & 0.04 & 0.05 & 0.05 \\ 
		Total & 0.05 & 0.13 & 0.21 & 0.36 & 0.56 & 0.79 & 1.0        
	\end{tabular}
\caption{Elemental abundance, $f_M^\alpha$, of the most relevant magnetic transition metals, $M$, among 
compounds presenting a relative error beyond a give threshold, $\alpha$ (in \%). The last row, labelled as `total', 
lists the fraction of compounds with errors exceeding $\alpha$ (e.g. 0.13 of the compounds have a predicted
$T_\mathrm{C}$ with error exceeding 30\%). The last column (error 0~\%) shows the elemental abundance over
the entire set.}
\label{table:2}
\end{table}

In general, as expected, we find that the elemental abundance grows as the error on the predicted $T_\mathrm{C}$
gets smaller. Such dependence is rather flat for Cr, Mn and Ni, mostly because a relatively limited number of 
compounds containing these three transition metals are found ferromagnetic above 300~K. The distribution of 
errors is thus dominated by Fe-containing, and partially Co-containing, ferromagnets, which are calculated with 
an accuracy comparable to that of the total set. We can then conclude that our best ML model is well balanced 
across chemical composition and does not favour any particular region of the chemical space.

\subsection{Ability of the model to extrapolate}
\begin{figure}[h!]
	\centering
	\includegraphics[width=\columnwidth]{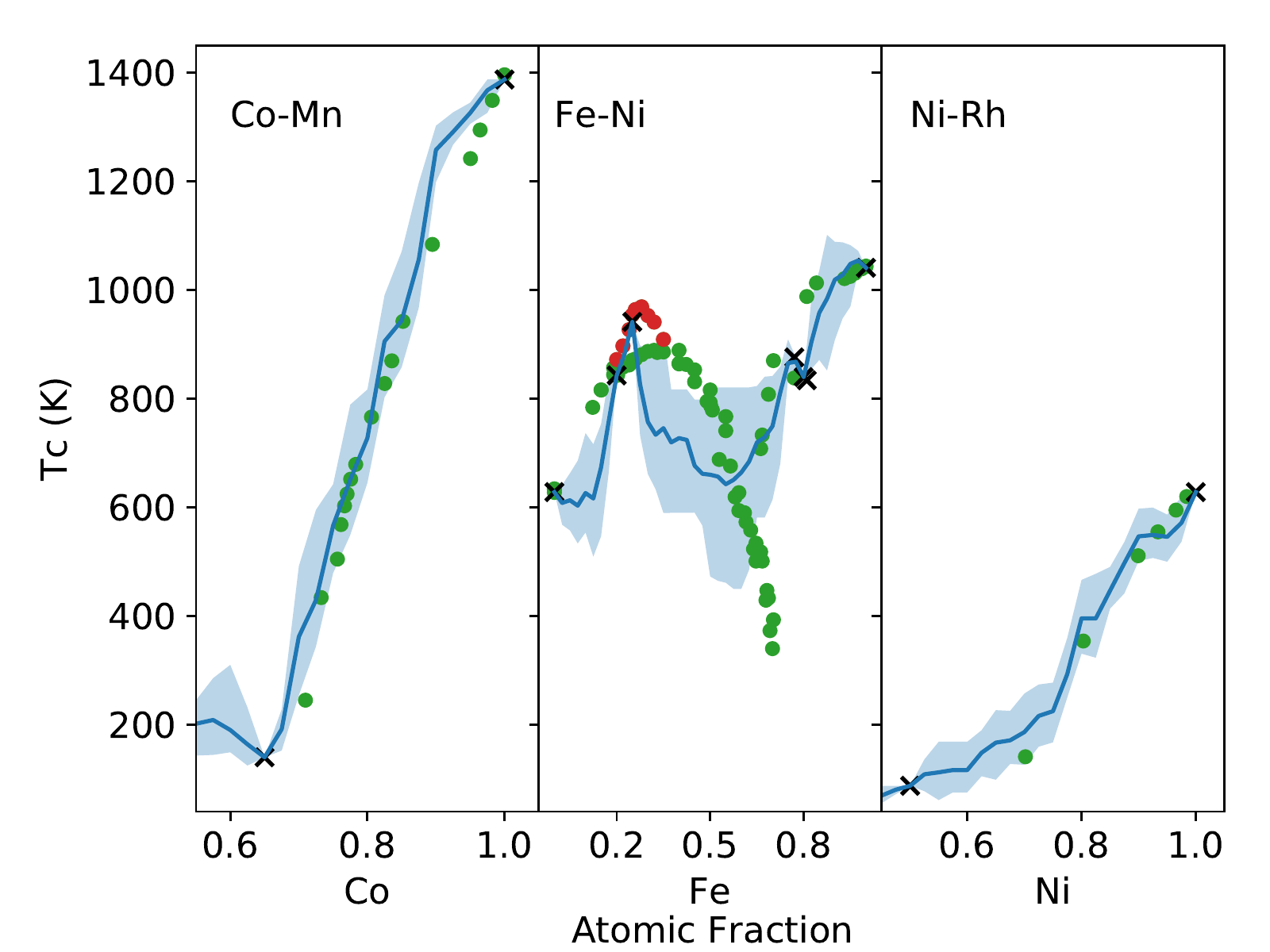}
	\caption{$T_\mathrm{C}$ prediction as a function of composition for three binary transition-metal systems, 
	namely Co-Mn, Fe-Ni and Ni-Rh. Data are presented as a function of the atomic fraction of one of the two
	species. The blue line traces the ML prediction, black crosses (green dots) are experimental 
	points included (not included) in the training set. The experimental results are taken 
	from references~\cite{menshikov1985magnetic, swartzendruber1991fe, crangle1960magnetization}. The 
	light-blue shadowed area corresponds to the range of predicted $T_\mathrm{C}$'s of subsets of the trees, 
	namely it indicates the uncertainty of the ML model. In the middle panel the red dots correspond to the 
	$T_\mathrm{C}$ associated to the FeNi$_3$ intermetallic phase, while the green ones correspond to 
	random Ni-Fe alloys.}
	\label{fig:phase}
\end{figure}

Next we demonstrate the ability of our model to extrapolate to regions of the chemical space where only a few
data points were present in the training set. Our first example consists in predicting the $T_\mathrm{C}$ as a function
of composition for three binary systems. In particular we consider Co-Mn, Fe-Ni and Ni-Rh for concentration
ranges where the alloys remain ferromagnetic. Our results are presented in Fig.~\ref{fig:phase}, where we show 
the prediction of our ML model against available experimental data. In particular we distinguish between the
data included in the training set (black crosses) and those that they were not (green dots). Using the Random 
Forest algorithm we can measure the models uncertainty by looking at the distribution of the predictions of the 
individual trees. In practice, we measure the uncertainty by grouping the trees into subsets of five trees each, 
averaging over these subsets and then using the minimum and maximum average as our uncertainty boundaries. 
In the test set 82~\% of the experimental values were within 50~K of the confidence interval. In the figure for 
every composition we present the average predicted $T_\mathrm{C}$ (blue line) and also the confidence interval
of the random forest algorithm (light-blue shadow).

Manganese and cobalt can form disordered alloys with a few possible crystal structures across the entire composition 
range~\cite{Ishida1990}. The magnetic phase diagram was determined sometime ago~\cite{menshikov1985magnetic}
and comprises both ferromagnetic and antiferromagnetic orders. In particular when the Co atomic fraction is
larger than about 0.68 the alloys are ferromagnetic with a $T_\mathrm{C}$ that monotonically increases as a 
function of the Co content. In contrast, Mn-rich alloys are antiferromagnetic with a N\'eel temperature that
this time monotonically increases as the Mn concentration gets larger. Our ML model predicts well the $T_\mathrm{C}$
for all the ferromagnetic phases, with a constant minimal error. For this case only two data points were included in 
the training set, namely elemental Co and the end-of-the-series alloy, Co$_{13}$Mn$_{7}$. 
Intriguingly, the model seems to be able to predict also the upturn in critical temperature occurring for Co atomic fractions 
lower than 0.68, although these correspond to antiferromagnetic phases. It is worth noting that the spread of values 
returned by the RF algorithm is certainly larger for these Mn-rich phases. 

The Ni-Fe system presents a different level of complexity, with the ferromagnetic order being present over the entire 
composition range~\cite{swartzendruber1991fe}. In the region of temperatures relevant for $T_\mathrm{C}$ the Fe-rich 
phases (for a Fe atomic fraction down to 0.65) are characterized by Ni-doped {\it bcc} iron with a $T_\mathrm{C}$ that grows 
monotonically as a function of the Fe fraction. In contrast, when the Fe atomic fraction is reduced below 0.65 the relevant 
phase is an {\it fcc} random alloy, which can be stabilized up to bulk Ni. In this case the $T_\mathrm{C}$ is non-monotonic, 
it increases with the Ni atomic fraction up to 0.70 (the maximum $T_\mathrm{C}$ is about 870~K for Fe$_{30}$Ni$_{70}$) 
and then it decreases down to the $T_\mathrm{C}$ of bulk Ni. Furthermore, there is also an ordered intermetallic FeNi$_3$ 
ferromagnetic phase, which persists over a relatively narrow composition range. This means that in Fe$_{1-x}$Ni$_{x}$
at $x\sim0.75$ there are two ferromagnetic phases with different $T_\mathrm{C}$'s. Also in this case our ML model
performs rather well. This time seven experimental data points were included in the training set, two for the elemental
Ni and Fe and five across the Fe$_{1-x}$Ni$_{x}$ alloys. In particular the composition corresponding to the maximum at the intermetallic FeNi$_3$ phase was included. Our
ML model interpolates well between these points, in particular in the Ni-rich part of the composition diagram. As 
expected from the fact that structural information are not included in the model, we are not able to distinguish the 
different $T_\mathrm{C}$'s associated to different structures.  

Finally, we look at the Ni-Rh binary system, an alloy where only one of the two elements is magnetic. As with several 
other elements of the Pt group, Rh is highly soluble in Ni and random alloys can be formed over almost the entire 
composition range. Here the $T_\mathrm{C}$ monotonically decreases from that of bulk Ni as Rh is added to the alloy. 
This continues up to a critical composition, found for a Ni atomic fraction of around 63\%, where the ferromagnetism 
disappears completely~\cite{PhysRevB.11.4552}. Our ML model is fully capable of describing such behaviour. In particular 
the ML model successfully predicts the critical concentration for the suppression of ferromagnetism at a Ni atomic fraction 
lower then 0.7. This is a rather compelling result. 

As a second example of the ability of our ML model to extrapolate to unexplored regions of the chemical space
we present the $T_\mathrm{C}$ diagram of the ternary system Al-Co-Fe.
\begin{figure*}[t]
	\centering
	\includegraphics[width=15cm]{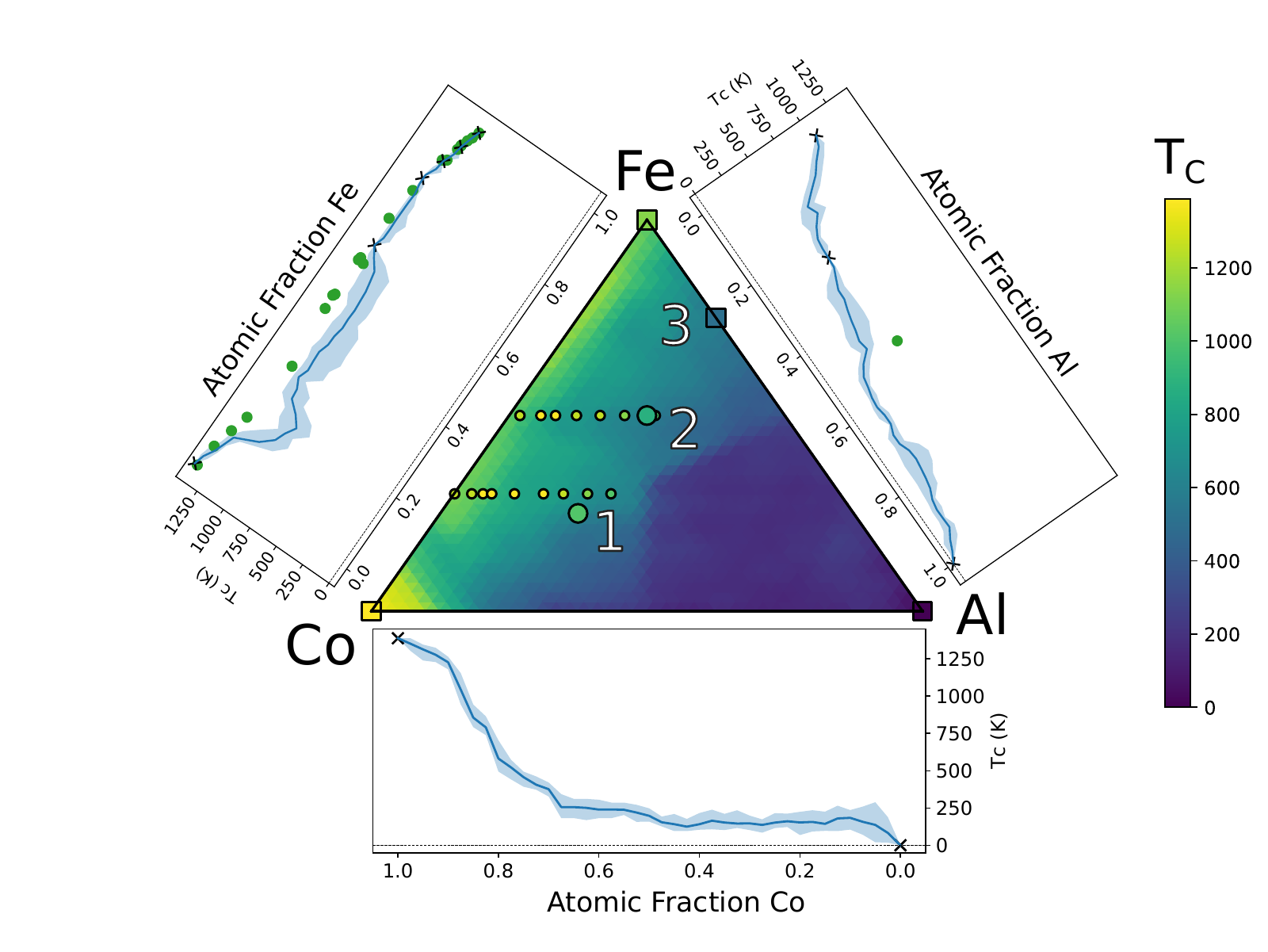}
	\caption{$T_\mathrm{C}$ prediction as a function of composition for the ternary system Al-Co-Fe. Data are 
	presented as a function of the atomic fraction of the three species and the $T_\mathrm{C}$ is expressed as
	a heat map. The figure also introduces a detailed analysis of the three relevant binary phase diagrams, where
	the blue line traces the ML prediction, black crosses (green dots) are experimental points included (not included) 
	in the training set. The light-blue shadowed area in the binary plots corresponds to the range of predicted 
	$T_\mathrm{C}$'s, namely it indicates the uncertainty of the ML model. The solid square (circles) included
	in the ternary $T_\mathrm{C}$ diagram are for experimental data included (not included) in the training
	set, with the colour code describing the $T_\mathrm{C}$. Numbers correspond to three known stoichiometric
	phases: 1) Co$_2$FeAl, $T_\mathrm{C}=$~1,000~K, 2) Fe$_2$CoAl, $T_\mathrm{C}>$~873~K, 
	3) Fe$_3$Al, $T_\mathrm{C}=$~573~K. The plot was partially made using python-ternary 
	\cite{harper2015python}.}
	\label{AlNiCo}
\end{figure*}
In Fig.~\ref{AlNiCo} we show the $T_\mathrm{C}$ across the ternary composition space as a colour-coded heat map
and that across the three possible binary systems as a standard graph, similar to those of Fig.~\ref{fig:phase}.
Also in this case for the binary systems we represent the data included in the training set as black crosses and those 
outside the training set as solid dots. The same convention is adopted for phases in the middle of the ternary diagram, 
where now the data included in the training set are solid square. Note that this ternary system includes three
stoichiometric magnetic phases, namely Co$_2$FeAl~\cite{Co2FeAl}, Fe$_2$CoAl~\cite{Fe2CoAl} and 
Fe$_3$Al~\cite{Fe3Al}. Furthermore, magnetic solid state solutions can be stabilized over a rather large 
composition space. 

As for the other binary systems investigated (see Fig.~\ref{fig:phase}) our machine learning model appears well
able to describe the main features of the three relevant binary alloys, namely Al-Co, Al-Fe and Co-Fe. This is
despite the fact that only a rather limited number of compounds across the various phase diagrams were included in 
our training set. For instance, the model is capable of describing the critical Al concentration for the appearance of
ferromagnetism in Al-Co~\cite{McAIieter1989}, which has been measured to be at an Al atomic fraction of about 
50\%. In this case only the end points of the composition diagram, namely elemental Co and Al, were present
in the training set, so that the ML model is able to extrapolate such critical concentration from the learning obtained
in other part of the chemical space.

The case of Al-Fe is relatively more complex. Our ML model is trained with only three data points from this binary 
system, namely the end points and the Fe$_3$Al stoichiometric phase. The resulting $T_\mathrm{C}$-versus-composition 
curve then predicts a monotonic reduction of the Curie temperature as the Al atomic fraction increases, with a 
rather smooth approach to $T_\mathrm{C}=0$ for pure Al. Experimental
data obtained for rf-sputtered thin films display a well-disordered {\it bcc} phase for Al atomic fractions up to 70\%,
followed by an amorphous phase between 75\% and 85\%, and then by an {\it fcc} phase for higher Al fractions. The 
disordered {\it bcc} phases are all ferromagnetic, while both the amorphous and the {\it fcc} ones are paramagnetic
down to 4.2~K~\cite{Shiga1985}. Although the precise position of such phase boundaries seems to depend 
somewhat on the details of the growth conditions~\cite{Sumiyama1990}, our ML model appears also in the case to be
able to capture such behaviour. 

At variance with the Al-Fe and Al-Co case, the $T_\mathrm{C}$ of the Co-Fe binary system presents a non-monotonic 
behaviour with composition. As we move from elemental Co to elemental Fe, first the $T_\mathrm{C}$ decreases with 
increasing the Fe content, up to a Fe atomic fraction of around 30\%. In this case the magnetic transition takes place within 
the Co-Fe $\gamma$-phase ({\it fcc} structure). Then, for Fe atomic fractions comprised between $\sim$30\% and 
$\sim$80\%, $T_\mathrm{C}$ first increases and then decreases, reaching up a maximum at around a 50-50 composition. 
Such behaviour effectively traces the phase boundary between the $\gamma$ and the $\alpha$ phases ({\it bcc} structure).
At the end of the composition range, for Fe atomic fractions above 80\%, $T_\mathrm{C}$ keeps decreasing down to that 
of elemental Fe, but the alloys remain in the $\alpha$ phase, namely the magnetic phase transition no longer traces the 
structural phase boundary \cite{Nishizawa1984}. 

Finally, we take a look at the $T_\mathrm{C}$'s for the ternary phases, namely in the center of the composition diagram.
In this case no experimental information was included in the training set. 
Two ternary stoichiometry compounds are known for Al-Co-Fe, namely the Heusler alloys Co$_2$FeAl and Fe$_2$CoAl. 
Their Curie temperature are 1,000~K for Co$_2$FeAl~\cite{Co2FeAl} and at least 873~K for Fe$_2$CoAl~\cite{Fe2CoAl}
($T_\mathrm{C}$ has not been measured with precision and only a lower bound is available). Our ML model predicts 
respectively 657~K and 580~K, hence it provides an underestimation of the real 
$T_\mathrm{C}$'s, although it ranks the materials in the right order. Additional experimental data are available 
across the Al-Co composition (for an Al atomic fractions not exceeding 30\%) and constant Fe atomic fractions
of 30\% and 50\%~\cite{Raghavan2005}. These data are reported as full circles in Fig.~\ref{AlNiCo} showing the ability
of our ML model to describe the general trend, namely a decrease in $T_\mathrm{C}$ as the Al atomic fraction increases. 
Also in this case some non-monotonicity is found in the experimental data for small Al concentrations, which is associated
to the fact that for such composition range the $T_\mathrm{C}$ traces the phase boundary between the $\alpha$ and
$\gamma$ phases. Our ML model appears to be able to trace such non-monotonicity.

\subsection{Incorporating structural information in the feature vector} \label{Incorporating Structure}
As constructed, our ML model does not include any information about the atomic structure of a given compound.
For this reason it is unable to distinguish the Curie temperatures of two polymorphs having the same chemical 
composition, a fact that has been discussed in connection to the Ni-rich part of the Fe-Ni $T_\mathrm{C}$ diagram 
(see Fig.~\ref{fig:phase}). We now present an attempt at overcoming such shortfall by including structural information 
in our feature vector. This is not a trivial task. 

Several strategies to encode structural information of materials in a way that satisfies the four criteria introduced in 
the Method section have been proposed. These include partial radial distribution functions \cite{schutt2014represent}, 
Voronoi tessellation \cite{ward2017including} and representations learnt by neural networks \cite{schutt2018schnet}. 
The issue with including structural information is that it massively increases the size of the input space, thus requiring 
much more training data to fully capture the space. This may not be a problem when data is abundant or can be easily 
generated, but it becomes one when the data is limited, as in our case. In fact, out of our entire dataset, only 792 
entires have an associated entry in the ICSD database~\cite{ICSD}. As such only this subset of data can be used in a 
ML model informed by structural parameters.

We have then constructed four different ML models, each one of them containing only a limited description of the
structural information of a compound. The first, denoted as \textit{Original + Volume}, associates to any compound,
in addition to the previous features, the unit cell volume per atom. The second one replaces the atomic fraction of 
each element present in a compound with its volume fraction, defined as $\mathbf{v}_\mathrm{chem}\cdot V$, 
where $V$ is the volume per atom of the material. This is denoted as \textit{Original with V-Frac}. The next two 
models, instead, include a more detailed representation of the atomic positions. We have taken inspiration from 
the work of Sch{\"u}tt et al. \cite{schutt2014represent}, who introduced structural information in the form of a 
radial distribution function,
\begin{equation}
f(r) = \frac{1}{N_\text{cell}} \sum_{i=1}^{N_\text{cell}}\sum_j \theta(d_{ij} - r) \theta(r+\Delta - d_{ij})\:.
\end{equation}
Here $N_\text{cell}$ is the number of atoms in the unit cell, the index $i$ runs over all the atoms in the unit cell, 
while the index $j$ over all the atoms neighbouring that at $i$ are within some distance cutoff, $\theta(x)$ is the 
Heaviside step function, $d_{ij}$ is the distance between the atoms $i$ and $j$ and finally $\Delta$ is the interval, 
a parameter that we must specify. Note that $\theta(d_{ij} - r) \theta(r+\Delta - d_{ij})$ is equal to one if 
$r < d_{ij} < r+ \Delta$ and vanishes otherwise. In our case we evaluate $f(r)$ at the points $r_n=n\Delta$, 
where $n$ is an integer, and the resulting function is added to the original feature vector ({\it Original + f} model). 

A second strategy, which tries to incorporate more information in the radial distribution function consists in 
defining an ``interaction-resolved radial distribution function''
\begin{equation}
f_\alpha(r) = \frac{1}{N_\text{cell}} \sum_{i=1}^{N_\text{cell}}\sum_j \theta(d_{ij} - r) \theta(r+\Delta - d_{ij})\delta_{\alpha, ij}\:.
\end{equation}
Here $\alpha$ represents a type of ``interaction'' and $\delta_{\alpha, ij}$ does not vanish, if the atoms at the positions
$i$ and $j$ define that given interaction. In practice, with ``TM'' meaning a transition metal atom, ``LA'' a lanthanide, and ``OT''
an atom of other type, we consider six types of interactions: TM-TM, TM-LA, TM-OT, LA-LA, LA-OT and OT-OT. 
Thus, the $f_\alpha(r)$ radial distribution function effectively defines the type-specific neighbourhood of a given atom.
Note that the two distribution functions are related to each other by the sum rule $f(r) = \sum_\alpha f_\alpha(r)$.
The rationale behind $f_\alpha$ is that the magnetic exchange interaction is, in general, 
dependent on the atom type, and determines the $T_\mathrm{C}$. As such, by containing the $f_\alpha$ our feature vector 
is expected to have some freedom to learn about the exchange interaction of a given compound. A similar, although more 
complex, representation was used also by Sch{\"u}tt et al.~\cite{schutt2014represent}. Here we have simplified the description 
to keep the dimension of the feature vector relatively low. This last model is denoted as {\it Original + $f_{ab}$}.

We then use 3-fold cross-validation on the 792-entires data subset with the Random Forests and KRR algorithm, 
and the results are shown in Table~\ref{table:3}. Note first that on this reduced data subset the original feature vector
(129-dimensional) is not able to generate ML models as accurate as before. For instance the $R^2$ coefficient of the 
RF algorithm is now only 0.75, compared with the previous value of 0.81. This is expected, since the pool of data
used for the construction of the model now has been drastically reduced. Unfortunately, we also find that any additional
feature added to the original vector generally makes the construction of an accurate ML model less successful 
with $R^2$ coefficients systematically smaller than those obtained for the original model regardless of the specific ML algorithm 
used. The exception here is when adding the volume as a feature. However, the improvement is so small that it is doubtful 
whether this is a significant result. There are two possible reasons, probably both at play, behind this result. Firstly, in all cases 
the feature vector dimension is larger, while the training dataset is smaller. Thus the models have now less data to train
but more information to use, effectively jeopardising their ability to learn. Secondly, we have now little control on the
balance of the data, meaning that we may have a disproportionated amount of information across the different regions of
the chemical space. 
\begin{table}[h!]
	\centering
	\begin{tabular}{l|c|c|c}
		Features & RF $R^2$ & KRR $R^2$ & Dim.\\
		\hline
		Original & 0.75 & 0.64 & 102 \\
		Original + Volume &  0.76 & 0.64 & 103\\
		Original with V-Frac & 0.75 & 0.61 & 129\\
		Original + $f$ & 0.74 & 0.63 & 122\\
		Original + $f_{ab}$ & 0.74 &  0.64 & 220\\
	\end{tabular}
	\caption{The 3-fold cross-validation $R^2$ score of the variously defined feature vectors for both Random Forests (RF) 
	and Kernel Ridge Regression (KRR). Note that here ``Original'' refers to the original set of features. The last column reports
	the dimension of the feature vector.}
	\label{table:3}
\end{table}

\section{Conclusion}

We have here described the construction of a ML model to predict the Curie temperature of ferromagnets solely
based on their chemical composition. The model has been entirely trained over available experimental data and
its construction did not involve any electronic structure calculations. We have considered several ML algorithms
and, by using Random Forest we have been able to generate a model presenting a mean absolute error of
only 57~K over a test set containing 767 compounds. Interestingly most of the error is associated to magnets 
with low $T_\mathrm{C}$, so that the model is capable to identify high-$T_\mathrm{C}$ ferromagnets. 

We have then made several attempts to include structural information into the description, but we have never been
able to outperform the models containing chemical data only. This is likely to be related to the smaller pool of data
to use for the training and to the larger dimension of our feature space. It is also interesting to note that ferromagnets,
being mostly metallic, are less sensitive than other magnetically ordered compounds to the structural details.
All in all our ML model can be viewed as a first rough guide to navigate the chemical space of magnetism, namely
as a first step toward a ML-driven magnetic materials discovery strategy.

\subsection*{Acknowledgment}

Financial support for this work has been provided by the Irish Research Council. We thanks Alessandro Lunghi
for useful discussions and for a critical read of the manuscript.

\bibliography{Paper}

\end{document}